\documentclass[twocolumn,nofootinbib,floatfix,10pt,aps]{revtex4-2}
\usepackage{amsmath}
\usepackage{amssymb}
\usepackage{wasysym}
\usepackage{graphicx}
\usepackage{color,soul}
\usepackage{physics}
\usepackage{siunitx}
\usepackage{dsfont}
\usepackage{float}
\usepackage{multirow}
\usepackage{mathdots} 
\usepackage[english]{babel}
\usepackage{blindtext}
\usepackage[english,nomargin,inline,marginclue,draft]{fixme}
\pdfpageheight\paperheight
\pdfpagewidth\paperwidth
\usepackage[colorlinks,linkcolor=blue,anchorcolor=blue,citecolor=blue,urlcolor=blue]{hyperref}
            
\fxusetheme{colorsig}
\FXRegisterAuthor{cg}{acg}{CG}  
\FXRegisterAuthor{th}{ath}{\color{blue}TH}  
\FXRegisterAuthor{ib}{aib}{\color{red}IB} 
\FXRegisterAuthor{sh}{ash}{\color{cyan}SH} 
\FXRegisterAuthor{db}{adb}{\color{green}DB} 
\FXRegisterAuthor{ps}{aps}{PS}
\makeatletter
\renewcommand*\FXLayoutInline[3]{%
  {\@fxuseface{inline}\ignorespaces{\color{fx#1}[#3: #2]}}}
\makeatother

\long\def\symbolfootnote[#1]#2{\begingroup%
\def\thefootnote{\fnsymbol{footnote}}\footnotetext[#1]{#2}\endgroup}

\def\nobreakbefore{%
  \relax\ifvmode\else
    \ifhmode
      \ifdim\lastskip > 0pt\relax
        \unskip\nobreakspace
      \else 
        \nobreakspace
      \fi
    \fi
  \fi
}
\let\oldcite\cite
\renewcommand\cite{\nobreakbefore\oldcite}

\begin{document}
\title{Floquet engineering Rydberg sub-THz frequency comb spectroscopy}

\author{Li-Hua Zhang$^{1,2}$}

\author{Zong-Kai Liu$^{1,2}$}

\author{Bang Liu$^{1,2}$}

\author{Qi-Feng Wang$^{1,2}$}

\author{Yu Ma$^{1,2}$}

\author{Tian-Yu Han$^{1,2}$}

\author{Zheng-Yuan Zhang$^{1,2}$}

\author{Han-Chao Chen$^{1,2}$}

\author{Shi-Yao Shao$^{1,2}$}

\author{Qing Li$^{1,2}$}

\author{Jun Zhang$^{1,2}$}

\author{Dong-Sheng Ding$^{1,2}$}

\email{dds@ustc.edu.cn}

\author{Bao-Sen Shi$^{1,2}$}
\affiliation{$^1$Key Laboratory of Quantum Information, University of Science and Technology of China, Hefei, Anhui 230026, China.}
\affiliation{$^2$Synergetic Innovation Center of Quantum Information and Quantum Physics, University of Science and Technology of China, Hefei, Anhui 230026, China.}
\affiliation{}
\date{\today}

\begin{abstract}
    Engineering a Terahertz (THz) frequency comb spectroscopy at atomic level advances the precisely measurement in spectroscopy and sensing. Current progresses on THz frequency comb rely on difference-frequency generation, optical parametric oscillation, and other methods. Generating a THz frequency comb poses challenges in source stability and achieving a narrow bandwidth, which traditional THz devices are difficult to achieve. Furthermore, accurately measuring the generated THz frequency comb necessitates a high-performance THz detector. Rydberg atoms are well-suited for electric field sensing due to their ultra-wide radio frequency transition energy levels, making them especially sensitive to external electric fields in the DC to THz bandwidth. However, there have been no reports about generating THz frequency comb spectroscopy at the atomic level until now. This work presents a THz frequency comb spectroscopy with Rydberg atoms, in which a Floquet comb-like transition is engineered through a time-periodic drive field. Our approach simplifies the setup required for THz frequency comb spectroscopy while extending the working bandwidth for Rydberg atomic sensors. The THz frequency comb spectroscopy at the atomic level reported in this article shows great potential for various applications in astronomy, remote sensing, spectral detection of biological samples, and other related fields.
\end{abstract}

\maketitle

\section{Introduction}

The Terahertz (THz) frequency comb has a wide range of applications in various fields, for example in application of spectroscopy, which can be used for analyzing chemical compounds, biomedical imaging, and biological molecules \cite{picque2019frequency,Yang:16,PhysRevX.4.021006}. There are several methods for generating THz frequency combs, including: frequency comb generated by quantum cascade lasers \cite{burghoff2014terahertz,lu2019room}, a passive mode-locking resonant-tunneling-diode oscillator \cite{hiraoka2022passive}, pump optical oscillators by femtosecond pulse laser\cite{hsieh2014spectrally} and frequency generation through a modulated graphene sheet \cite{PhysRevB.91.161403}.  

Rydberg atoms with a large dipole moment have exaggerated properties that make them especially fascinating for investigating their interaction with THz fields. At high THz field strengths, these atoms display nonlinear responses, which can be exploited for applications such as THz field measurement. This unique behavior of Rydberg atoms makes them promising candidates for studying the interaction between THz fields and matter, and also for advancing atomic level THz technologies. This THz frequency comb spectroscopy has several potential applications in THz signal measurement since the precision of the atomic level frequency comb could potentially measure the chemical vibrational and rotational states with high frequency resolution and accuracy, and this spectroscopy measurement also exhibits high spatial resolution as it is atomic level. This promotes perspective applications in fields such as chemical analysis, atmospheric monitoring, and astronomy.

Here, we have realized an atomic level THz (with central frequency larger than 100 GHz) transition frequency comb spectroscopy with an over 70\,MHz bandwidth in Rydberg atoms, in which the Floquet comb-like transition is engineered through a time-periodic driving THz field. In this process, the strong nonlinear interaction between Rydberg atoms and the input double-frequency bin THz field induces higher-order multi-photon transitions. Floquet engineering effectively enhances the probing bandwidth for sub-THz Rydberg atomic sensors by generating a series of Floquet sidebands (quasi-energy state) from the modulation frequency. As the drive field strength is increased, Floquet beat-note sideband transitions with an extraordinary range of over $\pm$\,9 orders are observed. This remarkable phenomenon significantly enhances the capabilities of the sensor. 

\begin{figure*}[htbp!]
\includegraphics[width=2\columnwidth]{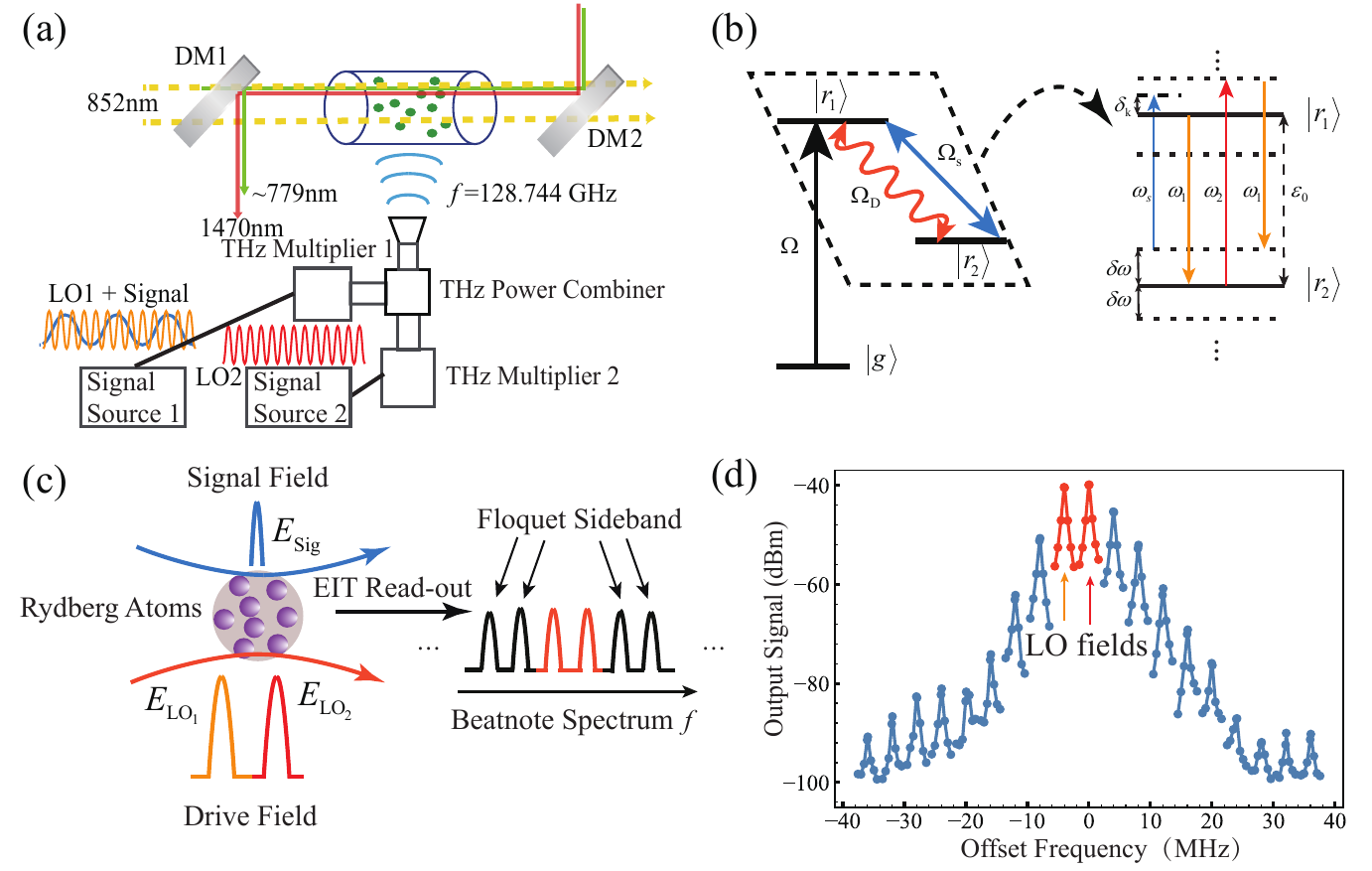}\caption{The energy level diagram of the experiment and the explanation for the extended beat transition peak for the Rydberg system under strong drive. For Fig.~\ref{fig1}(a), a strong equivalent bi-chromatic drive field is synthesized by combining two strong LO THz fields with a THz power combiner. At the same time, the weak signal field is generated by a two-tune modulation with attenuation on the LO1. (b) Under the strong drive field, a series of Floquet sidebands (quasi-energy levels) are generated in the two-level Rydberg system as shown in the dashed line box and the weak signal beat with the quasi-energy in Fig.~\ref{fig1}(d) The THz frequency comb spectroscopy measured in a range of $\pm$40\,MHz with a frequency interval of 4\,MHz for the THz field. }
\label{fig1}
\end{figure*}
            
\section{Setup and Model}

As shown in Fig.~\ref{fig1}(a), we use a three-photon scheme \cite{Carr:12} to excite and probe the Rydberg states ($6S_{1/2}\rightarrow 6P_{3/2}\rightarrow 7S_{1/2}\rightarrow 50P_{3/2}$). We scan the detuning of the coupling laser to obtain the electromagnetically-induced transparency (EIT) signal by collecting the probe beams with a photodetector. A pair of millimeter wave signals are generated by two vector signal sources (Ceyear, 1465F) and they are injected into the Frequency Multiplier Chain (FMC) ($\times$6) and ($\times$9) to generate the drive and signal THz fields. Then these two THz fields are combined through a THz power divider to form a bi-chromatic drive field. The frequencies of the two LO fields are set to 128.742\,GHz and 128.746\,GHz which is near-resonant with the Rydberg atomic transition, respectively, so their frequency difference is 4\,MHz in Fig~.\ref{fig1}(d). We demonstrate a Rydberg sub-THz frequency comb spectroscopy by Floquet engineering to extend the beat-note working range of the Rydberg atomic sensor in the sub-THz range. The Rydberg Floquet transition is driven through a strong effective bi-chromatic sub-THz field with a central frequency $\omega_{0}$ which is near-resonant with the Rydberg transition ($50P_{3/2}\longleftrightarrow 51D_{5/2}$) at 128.744\,GHz (Typically, 128.717\,GHz calculated by the python package ARC \cite{vsibalic2017arc}), then a series of Floquet states and a series of equivalent frequency bins are generated through the multi-photon dressing. Then the weak signal field couples (-23\,dB lower than the local field, typically) with these Floquet side bands and generates a beat signal together with the drive field. 

\begin{figure*}[htb]
\includegraphics[width=2\columnwidth]{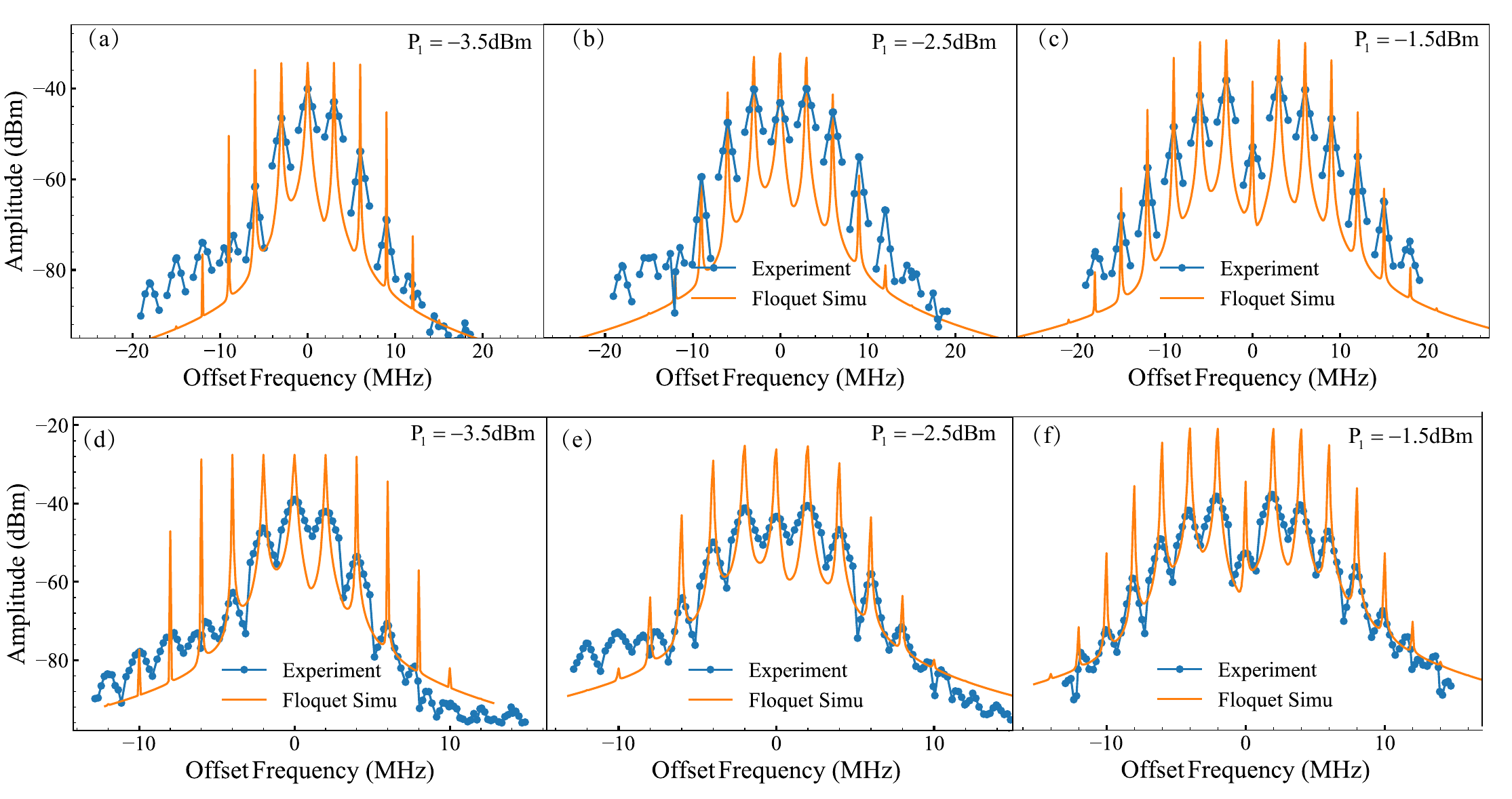}\caption{The frequency comb with varies LO power (For LO1 field, the power $\mathrm{P_{1}}$ output by the signal source 1 is changed from -4.0 to -1.5\,dBm ) and different modulation frequency (3\,MHz and 2\,MHz) is applied. (a)-(c): The frequency repetition frequency is set to 3\,MHz. (d)-(f) The frequency comb repetition frequency is 2\,MHz. The experimental (Blue dots) and the Floquet simulation (Orange curve) results are plotted and compared. The intensity of the output beat signal gets higher with the increase of the LO field power.}
\label{fig2}
\end{figure*}

The Floquet engineering provides a powerful toolbox for tuning the energy spectrum of the quantum system \cite{oliver2005mach,weitenberg2021tailoring}, realize non-trivial topological phase \cite{PhysRevB.82.235114,PhysRevLett.119.123601,mukherjee2017exp}, enhance the working bandwidth and the sensitivity of a quantum detector \cite{PhysRevX.5.041016,sciadv.abl8919,PhysRevLett.128.233201,PhysRevX.12.021061} and improving the coherent time of qubits \cite{PhysRevApplied.17.064006,PhysRevApplied.14.054033}.
In Fig.~\ref{fig1}(b), a simpl ified three-level system to show the detection protocol is illustrated. The three-photon Rydberg excitation in our experiment is eliminated to a single field probe and excitation with a Rabi frequency of $\Omega$. A strong Floquet drive field $\Omega_{\mathrm{D}}$ (Orange) and a weak signal field $\Omega_{\mathrm{s}}$ (Blue) couple with Rydberg energy levels $\ket{r_{1}}$ and $\ket{r_{2}}$. A series of Floquet quasi-energy levels appear under this strong drive field. The weak signal field probes these quasi-energy states through a multi-photon process. Different from the detection of the probe absorption spectra in other pump-probe methods \cite{PhysRevB.96.174518,PhysRevB.87.134505}, here the final output beat signal between the driven field and signal field is read out from the Rydberg EIT spectrum \cite{simons2019rydberg,jing2020atomic}.

We consider the two-level Rydberg states as a subsystem of the whole three-photon EIT system and we assume there is a weak effect for the drive field to the rest energy levels \cite{jiaorfrydberg,PhysRevA.58.2265}. Then we can reduce the multi-level model to a driven two-level subsystem depicted in Fig.~\ref{fig1}(b). In the experiment, the strong periodic drive field with a Rabi frequency $\Omega_\mathrm{D}$ consists of two LO fields with the same electric field intensity, and the Rabi frequency of each LO field is $\Omega$. The frequency of one of the LO fields (LO1) is $\omega_{1}$, the frequency of the other one (LO2) is $\omega_{2}$, and the central frequency of these two LO fields is $\omega_{0}$ ($\omega_{0}=[\omega_{1}+\omega_{2}]/2$). There is a small frequency difference $\Delta \omega=\omega_{1}-\omega_{2}$ ($\Delta\omega \ll \omega_{0}$) between them which equals the Floquet drive frequency. According to the sum and difference identities, the drive fields are separated into two parts, the high-frequency part oscillating with a frequency of $\omega_{0}$ couples the near-resonant Rydberg transition, and the low-frequency modulation part oscillating with a frequency of $\omega$ acts as a periodically driven field. We ignored the relative phase between the two components of the drive field, the expression for the total drive field Rabi frequency $\Omega_{\mathrm{D}}$ is written as
\begin{equation}
\begin{aligned}
\Omega_{\mathrm{D}} &=\Omega \cos (\omega_{1} t)+\Omega \cos [(\omega_{2}) t] \\
& \approx 2\Omega \cos (\Delta\omega t+\varphi / 2) \cos (\omega_{0} t+\varphi / 2).
\end{aligned}
\label{eq1}
\end{equation}
The anti-rotational wave term in the drive field Rabi frequency is also ignored. The expression of the drive field is given by
\begin{equation}
\begin{aligned}
\Omega_{\mathrm{D}}& \approx [ e^{i\omega_{0}t}+e^{-i\omega_{0}t}] \cos({\omega t}).
\end{aligned}
\label{eq1}
\end{equation}
\begin{figure*}[htp]
\includegraphics[width=2.05\columnwidth]{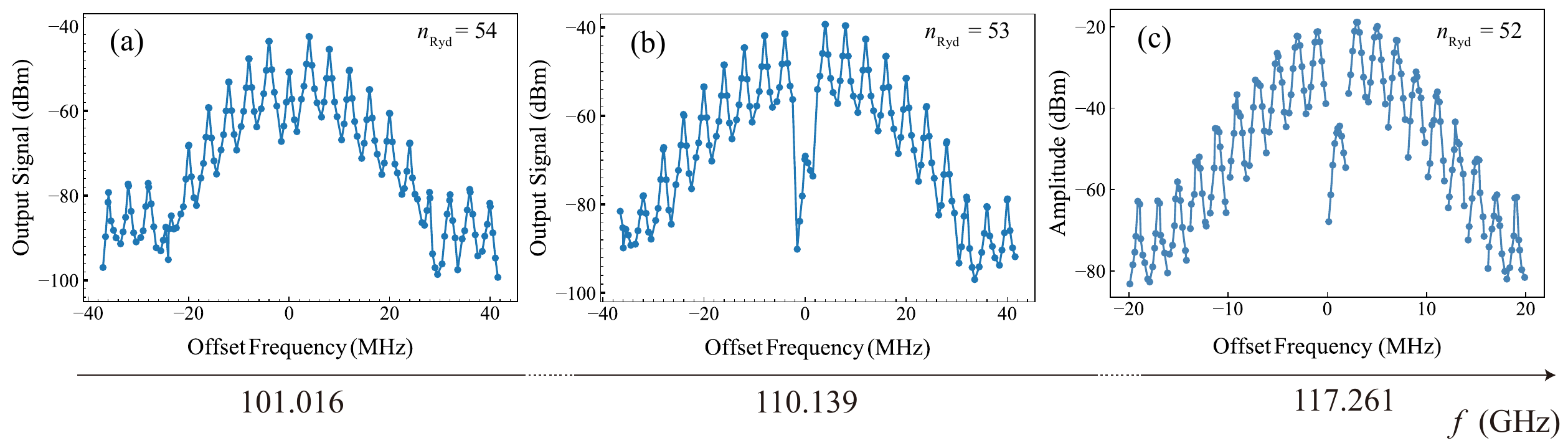}\caption{The Floquet beat comb spectroscopy with different principle quantum numbers $n_{\mathrm{Ryd}}$ is plotted. As we adjust the $n_{\mathrm{Ryd}}$ from 54 to 52, the central frequency of the comb spectra is tuned from 101.016\,GHz to 117.261\,GHz. }
\label{fig3}
\end{figure*}
Based on Eq.~\ref{eq1}, the subsystem Hamiltonian under rotation wave approximation is given by
\begin{small}
\begin{equation}
\hat{H}_{\mathrm{RWA_{0}}}=\left(\begin{array}{cc}
0 & {\begin{array}{c}\Omega e^{{i\omega_{0} t} }\cos(\delta  \omega t)\\+\Omega_{\mathrm{s}}\cos({\omega_{\mathrm{sig}}t})\end{array}} \\
\begin{array}{c}\Omega  e^{-i{\omega_{0} t}} \cos(  \delta\omega t) \\+\Omega_{\mathrm{s}} \cos({\omega_{\mathrm{sig}}t})\end{array} & \begin{array}{c}\varepsilon_{0}\end{array} 
\end{array}\right),
\end{equation}
\end{small}
where $\varepsilon_{0}$ is the Rydberg transition energy, $\Omega$ represents the Rabi frequency of each component of the bi-chromatic drive field and $\Omega_{\mathrm{s}}$ denotes the Rabi frequency of the signal field. 

We eliminate the high-frequency exponential oscillation term $e^{i\omega_{0}t}$ in the rotating frame, the subsystem Hamiltonian under the rotational wave approximation is given by
\begin{small}
\begin{equation}
\hat{H}_{\mathrm{RWA}_{1}}=\left(\begin{array}{cc}
0 & \Omega\cos( \Delta\omega t) +\Omega_{\mathrm{s}}\cos({\omega_{\mathrm{s}}t}) \\
\Omega \cos(\Delta\omega t) +\Omega_{\mathrm{s}} \cos({\omega_{\mathrm{s}}t})  & \Delta
\end{array}\right),
\label{eq3-flomfc}
\end{equation}
\end{small}
where the term $\omega_{\mathrm{off}}$ is the signal frequency offset relative to one of the drive field components so $\omega_{\mathrm{s}}=\omega_{\mathrm{sig}}-\omega_{0}$ and $\Delta$ denotes the detuning of the THz field to the Rydberg levels. By shifting the reference point of the Rydberg levels, we rewrite the subsystem Hamiltonian as 
\begin{small}
\begin{equation}
\hat{H}_{\mathrm{RWA}}=\left(\begin{array}{cc}
-\Delta/2 & \Omega\cos(\Delta\omega t) +\Omega_{\mathrm{s}}\cos({\omega_{\mathrm{s}}t}) \\
\Omega \cos(\Delta\omega t) +\Omega_{\mathrm{s}} \cos({\omega_{\mathrm{s}}t})  & \Delta/2
\end{array}\right).
\label{eq4-flomfc}
\end{equation}
\end{small}
According to the subsystem Hamiltonian denoted by Eq.~\ref{eq4-flomfc}, two electric field components of the bi-chromatic drive field are reduced to one time-periodic modulation term $\Omega \cos(\Delta\omega t)$. Under the bi-chromatic THz fields driving (Fig.~\ref{fig1}(b)), the multi-photon cycle transitions for the drive fields are shown. The LO1 field and the LO2 field are marked in red and orange arrows, respectively.

As shown in Fig.~\ref{fig1}(c) and (d), besides two zero-order beat resonance peaks which is due to beat-note between the signal field and the two LO fields directly, a series of Floquet beat sidebands appear on both sides with integer period of the drive frequency. The total beat process can be interpreted as a transition induced by the weak signal field between the Floquet states dressed by the bi-chromatic sub-THz field.
\begin{equation}
\left|r_1,n,-n-1\right\rangle \stackrel{\omega_s}{\leftrightarrow}\left|r_2, 0,0\right\rangle \text { or }\left|r_1,-n-1,n\right\rangle \stackrel{\omega_s}{\leftrightarrow}\left|r_2, 0,0\right\rangle
\label{eq3}
\end{equation}
If we use a three-mode representation, the transition process which contains the probe photon in the third mode is expressed as 
\begin{equation}
\begin{aligned}
\left|r_1,n,-n-1,+1\right\rangle &{\leftrightarrow}\left|r_2, 0,0,0\right\rangle \text {or}\left|r_1,-n-1,n,-1\right\rangle \\{\leftrightarrow}\left|r_2, 0,0,0\right\rangle
\label{eq4}
\end{aligned}
\end{equation}
where the index $r_{1}$,$r_{2}$ denotes the upper and lower Rydberg states, and $n$ photons of one frequency component (LO1 field) in the total drive field are emitted and $n$+1 photons of another component (LO2 field) are absorbed. At the same time, one signal THz photon is absorbed or emitted. Through this process, a series of additional side bands with a frequency of $ \omega_{1}(\omega_{2}) \pm n\, \Delta\omega$ are generated and the frequency of the beat signal observed in the experiment $\delta_{n}$ is given by 
\begin{equation}
\delta_{n}=\omega_{1}(\omega_{2})+n \cdot \Delta \omega-\omega_{\mathrm{sig}} (\delta_{n}<\Delta \omega, n=0,\pm 1, \pm 2 \ldots)
\label{eq4}
\end{equation}
This beat note corresponds with the phase factor of the exponential time-dependent oscillation term in the expansion of the subsystem Hamiltonian \ref{eq4}, which is determined by the order of the multi-photo transition process in Fig.~\ref{fig2}(b)

To observe the (beat) Floquet sidebands, we scan the frequency of the signal field and plot the read-out beat amplitude as the function of input frequency offset-$f_{\mathrm{off}}$ relative to the Rydberg resonance transition frequency (128.744\,GHz) in Fig.~\ref{fig1}(d). If the Rabi frequency $\Omega_{\mathrm{D}}$ of the driven field is relatively weak compared with the EIT linewidth, only two beat resonance peaks (orange) will exist at the central of Fig.~\ref{fig1}(d). We denote these two beat resonance peaks with 0-order side-band transition (when the signal field beats directly with two LO fields). With increasing the driven strength, the extra beat peaks become strong (blue) due to the Floquet dressing effect. When the Rabi frequency of the drive field is strong, the AC-stark shift of the strong drive field modifies the energy of the Floquet quasi-states (i.e., the signal field detuning from the quasi-states is changed.) The position of the beat note response does not change with the quasi-energy shifts because the beat note frequency between the drive and signal field is given by $\delta_{n}$  when the Floquet modulation frequency is maintained. 

To compare the theoretical beat response spectra with the experimental results, a more precise model considering the multi-mode (three-mode) Floquet-Born-Markov calculation is necessary since the two-mode Floquet calculation results in Fig.~\ref{fig1}(b) is symmetric about central peak and includes an apparent level shift which deviates from the resonance frequency. The amplitude of the output beat signal decreases when the signal field gets away from the Rydberg transition frequency. The two-mode Floquet calculation used here is similar to a single-mode case result, the signal field is just a weak probe \cite{PhysRev.138.B979,PhysRevB.87.134505}, the absorption and emission process containing one signal photon are considered in the Floquet calculation, so it does not influence the Rydberg energy levels.

As the output beat signal frequency in the Rydberg Floquet frequency comb spectroscopy is given by Eq.~\ref{eq4}, a series of beat resonance peaks appear at the integer multiple the bi-chromatic THz periodic drive frequency $\Delta \omega$, we can easily tune the comb repetition frequency by changing $\Delta \omega$. The numbers of the Floquet sideband and the covering range of the spectrum can be improved by increasing the drive field power. Here, we demonstrate the generation of frequency comb spectrum with different repetition rates (3\,MHz and 2\,MHz) as shown in Fig.~\ref{fig2}. In Fig.~\ref{fig1}(d), the repetition rate is set to 4\,MHz. Also, the central frequency of this frequency spectrum can be tuned by unitizing different Rydberg transitions and different THz carrier frequencies to adapt the application for different THz frequencies (up to 1\,THz) \cite{vsibalic2017arc} as shown in Fig~\ref{fig3}. Due to the limitations of the THz source we used, here, we demonstrate this between 90\,GHz and 140\,GHz.

Using a reduced two-mode Floquet theory \cite{PhysRevA.31.659}, we calculate the Floquet transition probability in this driven two-level subsystem model to simulate the output beat amplitude observed in our experiment. Here, we assume the beat amplitude observed in the EIT spectrum is proportional to the atom populations in the initial Rydberg state. This population is calculated according to the transition probability in the simplified two-level Rydberg subsystems without considering Doppler broadening, and power broadening \cite{PhysRevA.104.053101}. 

The three THz fields (two LO fields and the weak signal field) are reduced to two fields. 

The high-frequency part of the drive field is eliminated and the low-frequency beat between two LO fields is regarded as one drive field with a single time-dependent oscillation term. From the Floquet theory, we can convert the time-dependent Hamiltonian of the sub-system into a time-independent reduced two-mode Floquet Hamiltonian by expanding the former in a two-mode Floquet basis \cite{PhysRevB.87.134505,PhysRevA.79.032301,PhysRevX.12.021061}.
\begin{equation}
H(t)=\sum_{n,n^{\prime}=\pm 1} \sum_ {r=r_{1},r_{2}}
\mathrm{e}^{\mathrm{i}\left(n \Delta\omega+n^{\prime} \omega_{\mathrm{s}}\right) t} H^{[n,n^{\prime}]}_{r1,r2}.
\label{eq7}
\end{equation}
Here, $n$ and $n^{\prime}$ mean the Fourier index of the drive and signal field to mark different Floquet bases. If we compare the exponential phase factor in three mode Floquet representation before the Floquet state in Eq.~\ref{eq3} and the reduced ones in Eq.~\ref{eq7}, time-dependent phase factors of $e^{i(n\omega_{1}-(n+1)\omega_{2}+\omega_{\mathrm{sig}})t}$ or $e^{i(-(n+1)\omega_{1}+n\omega_{2}-\omega_{\mathrm{sig}})t}$ which corresponds to the oscillation frequency of the beat note signal observed in our experiment should appear before the Floquet state Eq.~\ref{eq7}. But in this reduced model, these two factors are $\mathrm{e}^{\mathrm{i}(n \Delta\omega+n^{\prime} \omega_{\mathrm{s}})}$. Based on the defination of the relative signal frequency ($\omega_{\mathrm{s}}=\omega_{\mathrm{sig}}-\omega_{0}$), the beat frequency between the signal THz field and the Floquet sidebands generated by the two LO THz fields (drive field) is expressed as
\begin{equation}
\delta_{n}=n \cdot \Delta \omega-\omega_{\mathrm{s}}=\omega_{1}(\omega_{2})+n \cdot \Delta \omega-\omega_{\mathrm{sig}} (\delta_{n}<\Delta \omega, n=0,\pm 1, \pm 2 \ldots)
\label{eq4}
\end{equation}
Two-mode Floquet matrix eigenvalue equation is given by
\begin{equation}
\mathbf{H}_{\mathrm{F}_2}|u\rangle=\epsilon|u\rangle
\end{equation}

The system time-dependent Hamiltonian is rewritten as 
\begin{equation}
\hat{H}(t)=\mathbf{H}^{[0]}+\mathbf{H}^{[ \pm 1]}\left(\mathrm{e}^{\mathrm{i} \Delta\omega t}+\mathrm{e}^{-\mathrm{i} \Delta\omega t}\right)+\mathbf{B}^{[ \pm 1,0]}\left(\mathrm{e}^{\mathrm{i} \omega_{\mathrm{s}} }+\mathrm{e}^{-\mathrm{i} \omega_{\mathrm{s}}t}\right)
\end{equation}
Assume the THz fields are near-resonant with Rydberg levels, the submatrices for the Floquet matrix are given by 
 $\mathbf{H}^{[0]}=\frac{1}{2}\left(\Delta \hat{\sigma}_z\right), \mathbf{H}^{[ \pm 1]}=\frac{1}{2}\Omega {\hat{\sigma}_x}$, and $\mathbf{B}^{[ \pm 1,0]}=\frac{1}{2}{\Omega _{s}} \hat{\sigma}_x$.
Then two mode Floquet Hamiltonian in the matrix form is

\begin{equation}
\mathbf{H}_{\mathrm{F_{2}}}=\left(\begin{array}{ccc}
\mathbf{H}_{\mathrm{F}}-\mathbb{I} \hbar \omega_{s} & \mathbf{B}^{[1]} & 0 \\
\mathbf{B}^{[-1]} & \mathbf{H}_{\mathrm{F}} & \mathbf{B}^{[1]} \\
0 & \mathbf{B}^{[-1]} & \mathbf{H}_{\mathrm{F}}+\mathbb{I} \hbar \omega_{s}
\end{array}\right)
\end{equation}

The single mode Floquet matrix 
\begin{equation}
\mathbf{H}_{\mathrm{F}_1}=\left(\begin{array}{ccccc}
\ddots & & \vdots & & \iddots \\
& \mathbf{H}^{[0]}-\mathbb{I} \hbar \Delta \omega & \mathbf{H}^{[1]} & 0 & \\
\cdots & \mathbf{H}^{[-1]} & \mathbf{H}^{[0]} & \mathbf{H}^{[1]} & \cdots \\
& 0 & \mathbf{~H}^{[-1]} & \mathbf{H}^{[0]}+\mathbb{I} \hbar \Delta \omega & \\
\iddots & & \vdots & & \ddots
\end{array}\right)
\label{eq8}
\end{equation}

By solving the eigenvalue problem, we can calculate the Floquet quasi-energy and the transition probabilities for the beat sidebands. The transition probabilities are expressed as \cite{PhysRevA.79.032301}.

\begin{equation}
\bar{P}_{{r_{1}} \rightarrow r_{2}}=\sum_{n,n^{\prime}} \sum_{r l_{1}l_{2}}\left|\left\langle { r_{1}, n,\pm 1 } | q_{r l_{1}l_{2}}\right\rangle\left\langle q_{r l_{1}l_{2}} | r_{2},0,0\right\rangle\right|^2
\end{equation}
Here, $\langle q_{r l_{1}l_{2}}|$ represents the Floquet eigenvector of the quasi-energy level. Since the intensity of the signal field is weak, only the lowest order process with one photon transition is taken into account in Equation \ref{eq8}. To obtain convergent calculation results of the Floquet frequency comb resonance peaks, 47 Floquet bases for $n$ and $n^{\prime}=0,\pm 1$ is used in our calculation. In Fig.~\ref{fig2}(a), the drive frequency is set as $\Delta \omega = 4\,\mathrm{MHz}$. The individual peak profile can be calculated by considering an instantaneous bandwidth model \cite{meyer2018digital,PhysRevApplied.18.014033}. When calculating the final frequency comb spectra, it is necessary to take into account the EIT linewidth, the system instantaneous bandwidth, and the photodetector bandwidth. The calculated Floquet frequency comb spectra need to be convoluted with an empirical response bandwidth function \cite{jiaorfrydberg}.

 \section{Discussion} 
In our experiment, we subtract the Floquet sideband by the beat signal from the Rydberg spectrum, which is directly achieved through a pump-probe scheme between two Rydberg energy levels \cite{PhysRevB.87.134505}; it is different from the scenarios using strong RF drive fields \cite{miller2016radio,bason2010enhanced}. By employing a simple modulation and detection setup, we not only obtain the amplitude, frequency, and phase information like the superhet scheme \cite{jing2020atomic} and the Rydberg microwave frequency comb spectrometer \cite{PhysRevApplied.18.014033}, but also realize a Floquet frequency comb detector with a broad and real-time probe range in the sub-THz band. The development of an atomic THz frequency comb holds great potential for advancing various fields and paves the way for technological innovation in the realm of atomic THz. In this experiment, the stability of the sub-THz frequency comb primarily relies on the stability of the THz field source. However, the high-lying Rydberg states used in generating the THz frequency combs are sensitive to disturbances caused by stray electric and magnetic fields, laser noise, and fluctuations in the surrounding gas pressure. 

The main difference between optical frequency comb spectroscopy and atomic THz frequency comb spectroscopy lies in their respective operating frequency ranges. Optical frequency comb spectroscopy involves generating a dense set of light waves with evenly spaced frequencies, typically used for precise measurement of optical frequencies. In contrast, atomic THz frequency comb spectroscopy generates a series of atomic transition combs with evenly spaced frequencies in the THz frequency range. This technique is mainly used for spectroscopy measurements in the THz wave band. Unlike optical frequency comb spectroscopy or observing the EIT optical spectrum under a strong drive field through scanning the laser frequency, the THz frequency comb spectroscopy demonstrated here relies on the beat phenomena occurring between the Floquet sidebands and a scanned signal field. 

The detection method in an atomic THz frequency comb is also different from the traditional optical frequency comb. People measure the frequency of laser beam by combining it with an optical frequency comb. The resulting beat signal is then detected using a photodetector and analyzed using a spectrum analyzer. By comparing the beat frequency with the known frequency of the frequency comb lines, the frequency of the laser beam can be accurately determined. It provides a highly accurate and stable method for measuring the frequency of lasers and other optical sources. For the atomic THz frequency comb reported here, we can combine a THz source with an atomic THz frequency comb in an atomic system. This process generates new frequencies, which creates a beat frequency in the probe field. There are beat signals from the mixing process between the THz field and the individual electric field components of the THz frequency transition comb lines in a Rydberg system, thus the frequency of the THz signal can be accurately determined. As our experiment is conducted in a free-space scheme, future improvements by using a THz cavity or waveguide to increase the coupling strength between the THz and the Rydberg atoms can help to improve the bandwidth and intensity of the spectroscopy. Benefiting from the wide-band electric response of the Rydberg atoms and the highly tunable comb repetition rates for this Floquet comb generation method, a wide-band Rydberg atomic THz frequency meteorology standard can be constructed with more effort.

 \section{Conclusion} 

In summary, we have observed a Rydberg sub-THz frequency comb spectroscopy, in which a bi-chromatic time-periodical THz field is used to drive the Rydberg energy levels and a beat Floquet side-bands (frequency comb spectrum) of $\pm 9$ orders are observed on Rydberg EIT spectrum. The frequency separation of this frequency comb spectrum can be easily tuned by changing the THz modulation frequency, thus expanding the range of working bandwidth for the Rydberg THz sensor. Although this protocol can also be applied to the microwave range, the performance may be affected by the limited Rydberg linewidth. The (near) resonant coupling between the drive field and the Rydberg energy levels provides a strong Floquet dressing on the Rydberg atom system, and the beat sideband read-out offers an alternative method to acquire the quasi-energy structure of the Rydberg atoms. By selecting different spectral lines in the THz frequency range, it is possible to realize a higher-frequency THz frequency comb and expand the overall bandwidth. This unique THz frequency comb spectroscopy technique offers a promising approach for precise THz spectroscopy, sensing, and metrology at the atomic level.

\section{Declarations}

\subsection{Ethical Approval and Consent to participate}
Not applicable
\subsection{Consent for publication}
Not applicable
\subsection{Availability of supporting data}
The datasets used and/or analysed during the current study are available from the corresponding author on reasonable request.
\subsection{Competing interests}
The authors declare that they have no competing interests.
\subsection{Authors' contributions}
D-S.D. and L-H.Z conceived the idea. L-H.Z implemented the physical experiments with  Z-K. L., Q-F.W., B.L, S-Y.S., Z-Y.Z. L-H.Z, T-Y.H, H-C.C, J.Z., and Q.L. built the theoretical model. D-S.D. and L-H.Z
wrote the manuscript. All authors contributed to discussions regarding the results and analysis contained in the
manuscript. D-S.D., B-S.S., G-C.G. support this project.
\subsection{Funding}
This research is funded by the National Key R\&D Program of China (Grant No. 2022YFA1404002), the National Natural Science Foundation of China (Grant Nos. U20A20218, 61525504, and 61435011), the Anhui Initiative in Quantum Information Technologies (Grant No. AHY020200), the major science and technology projects in Anhui Province (Grant No. 202203a13010001) and the Youth Innovation Promotion Association of the Chinese Academy of Sciences (Grant No. 2018490).
\bibliography{article}
\end{document}